\documentclass{nature}

\usepackage{amssymb}
\usepackage{graphics,graphicx}
\usepackage[super,comma]{natbib}
\usepackage{lineno}

\newcommand\gr{$\gamma$-ray}
\newcommand\grs{$\gamma$-rays}

\newcommand{\sn}{SN~2022jli}
\newcommand{\gsrc}{Gsrc}

\title{An extragalactic gamma-ray binary formed in supernova 2022jli}

\author{Pengfei Zhang$^{1,*}$, Zhongxiang Wang$^{1,*}$, Shunhao Ji$^{1}$}

\begin{document}


\maketitle

\begin{affiliations}
\item Department of Astronomy, School of Physics and Astronomy, Key Laboratory 
of Astroparticle Physics of Yunnan Province, Yunnan University, Kunming 650091, 
China

\end{affiliations}

\begin{abstract}
	On May 5 2022, a type Ic supernova (SN) explosion SN~2022jli was
	discovered. This SN showed additional optical emissions, which were
	found to exhibit 12.4-day periodic undulations and concordant 
	periodic
	velocity shifts. These key features likely indicate a compact object
	in a binary system was formed. A faint $\gamma$-ray source
	was also detected at the position of the SN and upon checking
	the $\gamma$-ray photons' arrival times, it was revealed that
	the same 12.4-day periodicity was likely present.
	Here we report our detailed
	analysis results for the $\gamma$-ray source. Not only was the 
	$\gamma$-ray emission detectable for $\sim$1.5\,years since the
	discovery time, but a strong modulation at period 12.5\,day
	was also clearly determined. Considering the newly formed compact 
	object to be a neutron star or a stellar-mass black hole, the 
	putative binary, having an orbital period of 12.5\,day,
	is likely the first extragalactic high-energy system detected.
	The system may serve as a valuable example for
	the formation of many such binaries observed in the Milky
	Way and nearby galaxies.
\end{abstract}

\section*{Introduction}

Stars, especially massive ones, are mostly found to be formed in binary or 
multi-object systems \cite{md17,tv23}. It has been realized that because of 
the binarity, the majority of massive stars will evolve as a result 
of their interactions with their companions \cite{san+12}.
Depending on initial conditions such as the orbital period and the masses of a 
binary, it will go through different series of evolutionary 
processes, ending up as one of different types of products \cite{tv23}. 
One such typical product is the X-ray binaries (XRBs); 
for example, a binary consisting of 20\,$M_{\odot}$ plus 8\,$M_{\odot}$ 
stars with a several-day orbital period will evolve into a neutron star (NS) 
XRB with a high-mass supergiant companion \cite{tv23}. 
Nearly 500 XRBs have been found within the Milky 
Way \cite{for+23,neu+23,for+24} and more populations of XRBs have also been 
identified in our local group of galaxies and beyond \cite{lvv05,lvv07,gil+22}.

XRBs are significant X-ray sources that are powered by accretion onto
compact-star primaries, either NSs or black holes (BHs), from 
their companions. Among those in the Milky Way, less than
10 have been further classified as gamma-ray binaries (GRBs) because
the energy spectra of these binaries peak at high-energy \grs\
(energies $\gtrsim$ 100\,MeV) \cite{dub13,cm20}. In addition,
XRBs with relativistic collimated jets, known as
microquasars, are also able to emit \grs. It has been understood that when 
the compact
star is either a pulsar or a BH with jets, significant \gr\ emissions can be
produced in processes involving high-energy particles provided by either 
the pulsar's wind or the BH's jets, respectively \cite{dub13}.

SN~2022jli was discovered on 5 May 2022. It was a type Ic supernova (SN)
occurring in the nearby galaxy NGC~157 at a distance of 22.5 megaparsec 
(Mpc) \cite{che+24}. Its optical emission was a double-peaked type;
in the declining multiband optical light curves after the second peak, 
12.4-day periodic undulations were observed, and concordant periodic
velocity shifts were detected in its narrow H$\alpha$ emission. These
features point to the likely formation of a compact remnant, the result of
the SN explosion of a massive progenitor, in a binary system with 
orbital period $P_{\rm orb} \simeq 12.4$\,day \cite{che+24}. In addition, 
associated \gr\ emissions were reportedly detected
mainly between Sept.--Dec. 2022, and a detailed check of the arrival times of
1--3\,GeV photons from the direction of \sn\ hinted that the same 12.4-day
periodicity was present in the \gr\ emission as well.

We have conducted detailed analyses of the \gr\ data obtained with the Large 
Area Telescope (LAT) onboard the Fermi Gamma-ray Telescope (Fermi). 
We report here the result, which is that there was a clear periodic 
modulation in the one 
year since the discovery time (MJD~59704.17; hereafter set as $T_0$) of 
\sn. Combining the results reported in ref.~\citenum{che+24}, this binary
is likely the first extragalactic GRB observed.

\begin{figure}
\begin{center}
\includegraphics[width=0.49\linewidth]{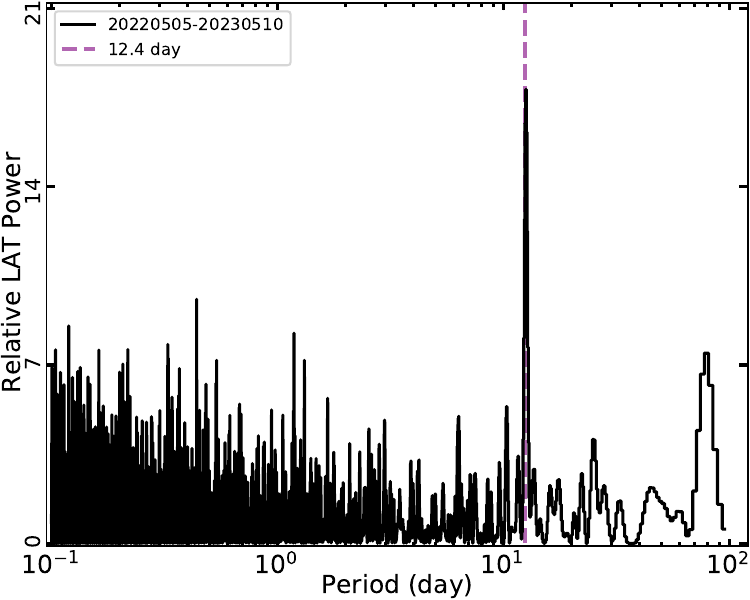}
\includegraphics[width=0.49\linewidth]{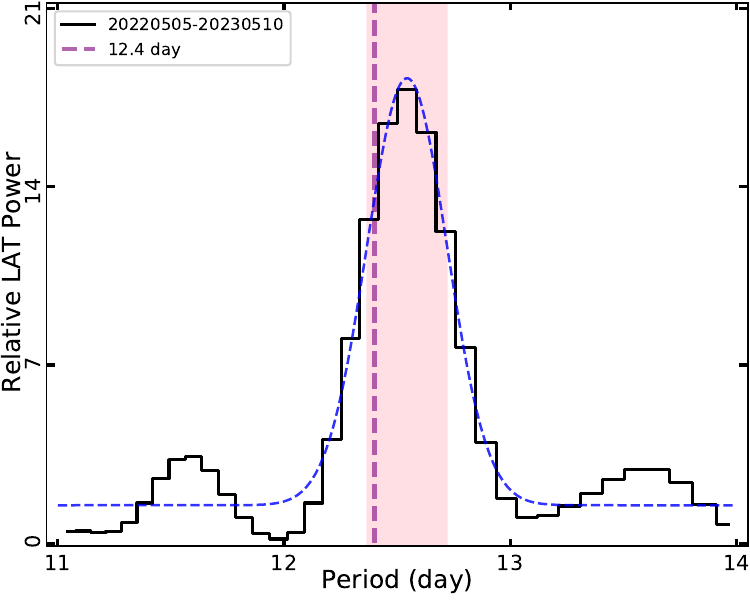}
\end{center}
	\caption{{\it Left:} LSP power spectrum of \gsrc\ obtained from the 
	LAT data from May 5 2022 to May 10 2023. A periodic signal near 
	12.4\,day is clearly visible. {\it Right:} LSP power spectrum of \gsrc\
	in a frequency window of from 1/14.0 to 1/11.0\,day$^{-1}$.
	A Gaussian fit (blue dashed line) was used to 
	determine the period and its uncertainty $P = 12.54\pm 0.18$\,day
	(marked by the pink region). In both panels, the 12.4-day value is 
	marked with a purple dashed line.
	\label{fig:lsp}}
\end{figure}

\section*{Results}
\subsection{Signal at period 12.54 day}

We analyzed the Fermi LAT data for \sn\ and found that the \gr\ emission 
was likely detectable from $T_0$ to Oct. 27 2023 (MJD~60244; see Methods
section `Fermi \gr\ data and source model'), much longer than that previously
reported in ref.~\citenum{che+24}.
By applying a modified aperture photometry method to the data in that time
period, we constructed a time series of 500\,s binned counts, which were
summed from probability-weighted \gr\ events in each time bin.
The events were in the energy range of 0.1--500\,GeV, selected in 
a 6$^{\circ}$ radius aperture with the center placed at the \gr\ counterpart
(hereafter named as \gsrc) to \sn\ (see Methods section 
`Periodicity determination'). Applying the Lomb-Scargle Periodogram
(LSP) \cite{l76,s82,zk09} analysis to the time series, the resulting
power spectrum clearly shows a periodic signal
near the reported 12.4-day period \cite{che+24}, which
was found to have a
maximum power value of 17.8 in the time period of T$_0$ to
approximately May 10 2023 (MJD~60074; left panel of Fig.~\ref{fig:lsp}). 
By fitting the power peak with a Gaussian function
(right panel of Fig.~\ref{fig:lsp}),
we determined the period $P = 12.54\pm 0.18$\,day.
This signal was estimated to be at a confidence level of 5.3$\sigma$
(see Methods section `Periodicity determination').

\subsection{Phase-resolved properties of the \gr\ emission from \sn}
\begin{figure}
\centering
\includegraphics[width=0.48\linewidth]{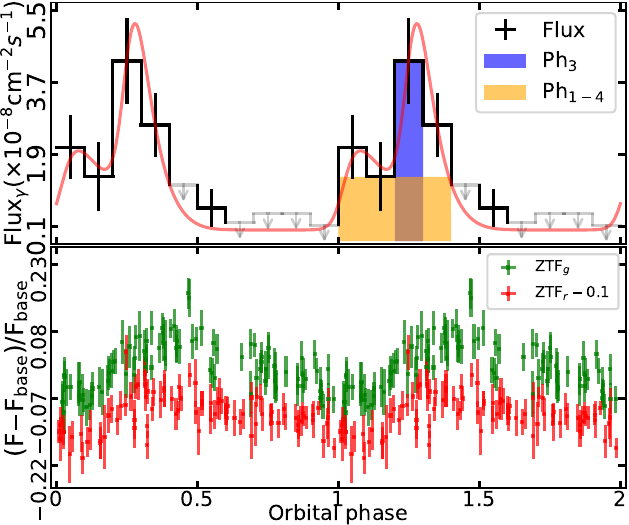}
\includegraphics[width=0.50\linewidth]{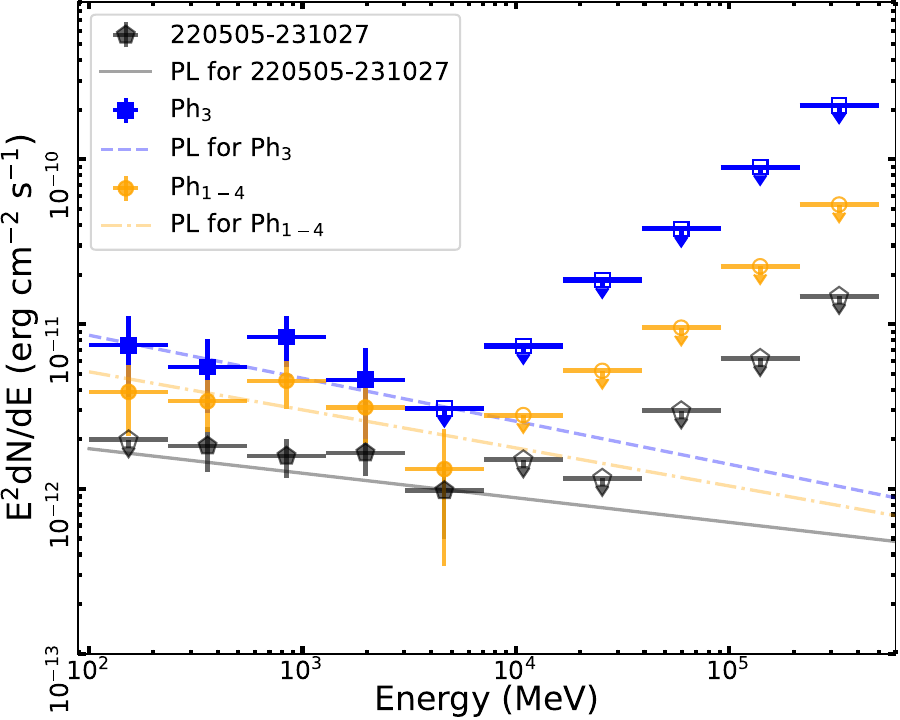}
	\caption{{\it Left:} folded \gr\ light curve ({\it top}) and 
	ZTF optical light curves ({\it bottom}). 
	A model fit, consisting of a jet plus a counter-jet 
	with the jets' height being 1.6$\times 10^{12}$\,cm
	(see ref.~\citenum{dch10} for details), to the \gr\
	light curve is shown (red solid line).
	{\it Right:} spectra obtained from the data in Ph$_3$ (blue) and
	Ph$_{1-4}$ (orange), where the phase ranges are marked blue and 
	orange respectively in
	the left top panel. For comparison, the spectrum from the data
	in the whole detectable time period (i.e., from
	May 5 2022 to Oct. 27 2023) is also shown. 
	The PL fits to the spectra are respectively plotted as the blue dashed 
	line, orange dash-dotted line, and black solid line. 
\label{fig:plc}}
\end{figure}

Using this period, which is consistent with the 12.4-day period
within the uncertainty,
a folded \gr\ light curve whose phase zero was set at $T_0$
was constructed (left panel of 
Fig.~\ref{fig:plc}; see Methods section `Phase-resolved analysis').
The light curve shows that
the \gr\ emission was detected only in phase 0.0--0.4, which matches
the rising part of the optical undulations. Phase-resolved spectra 
from phase 0.0--0.4 and phase 0.2--0.3 were obtained, where the latter phase 
range is that of the modulation peak. The spectra were detected
in an energy range 
of $\sim$0.1--3\,GeV, described with a power law (PL) with photon 
indices of $\simeq$2.2.
Comparing the spectra with that obtained from the data of
the whole detectable time period (i.e., from $T_0$ to MJD~60244), 
no significant spectral differences could be determined.

\section*{Discussion}
It has been extensively discussed for every aspect of \sn\ that this SN was 
likely an explosion in a binary system \cite{che+24}. The key features are the 
periodic undulations seen in optical light curves and concordant periodic 
velocity shifts seen in the narrow H$\alpha$ emission line. The 
emissions related to the second peak of the optical light curves
were induced by accretion onto the compact 
star that was newly formed in the SN, thus revealing the
orbital periodicity, and the narrow H$\alpha$ line was due to the hydrogen 
material
contained in the accreted mass from the envelope of the companion star. 
While deep X-ray observations conducted at the end of the optical light curves
only resulted in non-detection upper limits, the non-detection could be
because of the large optical depth caused by the surrounding ejecta of
the SN.

According to the orbital period distributions of Galactic XRBs,
a $P_{\rm orb} \simeq 12.5$\,day system would be more likely to be
a so-called high-mass XRB (HMXB) \cite{for+23,for+24} if the companion of 
the binary in \sn\ is
a high-mass, $\sim$10\,$M_{\odot}$ star that will have strong mass loss  
either through its wind or by overfilling its Roche lobe in the subsequent
evolutionary years.
Indeed, in order to match the observed H$\alpha$ velocity shifts, the possible
mass of the companion could be, for example, 5 or 15\,$M_{\odot}$ for a NS 
primary,
or 50\,$M_{\odot}$ for a 10\,$M_{\odot}$ BH primary (see Extended Data Fig. 10
in ref.~\citenum{che+24}). When the companion components of XRBs are 
considered, Galactic GRBs belong to the HMXB class since their companions are 
$\sim$10--30\,$M_{\odot}$ spectral-type O/B stars \cite{dub13, cm20}. 
In this regard, \sn's binary would be the first extragalactic GRB observed. 

The \gr\ emission of this putative binary was described with a PL, and its
folded light curve was highly modulated.
These properties
are different from those of Galactic GRBs as their \gr\ emissions
are similar to that of a pulsar, often described with a PL with an exponential 
cutoff at several GeV
energies \cite{dub13}; it has been tentatively argued that all the GRBs 
have a pulsar primary. Instead, the \gr\ properties are
reminiscent of those of the Galactic microquasars, namely 
Cygnus~X-1 \cite{mzc13,zdz+17} and Cygnus~X-3 \cite{cyg09,zdz+18}. These two 
XRBs were observed to have \gr\ 
emissions during their hard state when jets were present; the emissions were
well fitted with a PL, and those of Cygnus~X-3 were highly modulated on
its orbital period.

The \gr\ emissions of Cygnus~X-3 are well explained with a model that
considers
anisotropic Compton scattering of blackbody photons from the companion by
the relativistic electrons of its jets \cite{dch10,zdz+18}. We tested to 
use this model
to fit the folded light curve, where we adopted the BH binary system that
can provide a match to the velocity shifts (i.e., 10\,$M_{\odot}$ BH, 
50\,$M_{\odot}$ companion, orbital eccentricity $e = 0.8$, and inclination angle
of the binary $i = 90^{\circ}$; ref.~\citenum{che+24}).
The radius ($\sim 8\times 10^{11}$\,cm) and effective temperature 
($\sim 5\times 10^{4}$\,K) of the companion were estimated from the relations
given in ref.~\citenum{eke+15}. Because there are a few parameters in 
the model (in which a jet plus a counter-jet are included; see details in
ref.~\citenum{dch10}) and
limited information is known for the putative binary, we fixed the bulk 
velocity
of the jets at 0.1$c$ (where $c$ is the light speed) and the polar angle
of the jets at 60$^{\circ}$ (note that $0^{\circ}$ is when the jets are 
perpendicular to the orbital plane), and found that the jets with 
a height of $\sim 1.6\times 10^{12}$\,cm can provide acceptable
fits to the folded light curve (left panel of Fig.~\ref{fig:plc}). 
It is interesting to note that the BH binary system we adopted has a 
separation distance of
1.2$\times 10^{12}$\,cm at periastron, which is comparable to the jets' size.
This is similar to that found in the microquasars \cite{dch10,dub13}.

The detection of the \gr\ emission highly modulated on the optical-undulation
period greatly strengthens the likelihood of a compact-star binary
forming in \sn. At a distance of 22.5\,Mpc, if the compact star is a NS or 
a stellar-mass BH,
it would have experienced super-Eddington accretion in
order to power the observed optical undulation emission \cite{che+24}. The
apparent \gr\ luminosity of this putative binary (Table~\ref{tab:res}) was also
$>$1000 times the Eddington limit (for a stellar-mass compact star), 
potentially making it an ultraluminous \gr\
source, in parallel to the ultraluminous X-ray sources detected in the
X-ray sky \cite{pw23}.

\begin{methods}

\subsection{Fermi \gr\ data and source model}

The Large Area Telescope (LAT) onboard the Fermi Gamma-ray Space Telescope
(Fermi) has been scanning the entire sky since Aug. 2008.
We selected LAT's Pass 8 Front+Back events (evclass = 128, evtype = 3) in 
the 0.1--500~GeV energy range from Aug. 4 2008 to Jul. 7 2025 
(MJD~54682.687--60863.123).
A $20^{\circ}\times20^{\circ}$ region of interest (RoI) was set with the center
placed
at the position of the new \gr\ source \cite{che+24} (i.e., Gsrc),
R.A. = 8$^\circ.620$, Decl. = $-$8$^\circ.425$ (J2000.0). 
To minimize \gr~contamination from the Earth's limb, we excluded events 
with zenith angle $>90^{\circ}$. High-quality events within good time 
intervals were selected by applying the filter expression 
``DATA\_QUAL$>$0~\&\&~LAT\_CONFIG==1." The ``P8R3\_SOURCE\_V3"
instrument response function and Fermitools~2.2.0 were used in our analyses.

A source model was built based on the Fourth Fermi Gamma-ray LAT (4FGL)
catalog Data Release 4 (4FGL-DR4) \cite{4fgl-dr4}, with \gsrc\ added in as
a point source, by running the Python script make4FGLxml.py.
A power-law (PL) spectral form, $dN/dE=N_0(E/E_0)^{-\Gamma}$, where
$\Gamma$ is the PL index, was used for the emission of \gsrc.
The Galactic and extragalactic isotropic components,
files gll\_iem\_v07.fits and iso\_P8R3\_SOURCE\_V2\_v1.txt respectively, 
were also included in the source model. In our following
analyses, the normalizations of these two background components were
always set as free parameters.

The data in our analyses were from $\sim$17\,years of LAT observations,
longer than the 14 years of the data used to build 4FGL-DR4. We thus 
	updated the parameters in the source 
model by performing binned maximum likelihood analysis. 
In this analysis, flux normalizations and spectral parameters of
	sources within 5$^\circ$ of \gsrc\ were set free,
flux normalizations of sources within 5$^\circ$--10$^\circ$ and variable 
sources $>$10$^\circ$ of \gsrc\ were also set free,
	and all other parameters were fixed at values given in 4FGL-DR4.
From this analysis, we obtained a TS value of 22.8 for \gsrc;
the best-fit parameters were $\Gamma = 2.51\pm 0.20$ and photon flux 
$F_{\rm ph}=2.8 \pm 1.3\times10^{-9}$~photons~cm$^{-2}$~s$^{-1}$ 
in 0.1--500~GeV.
These values are also provided in Table~\ref{tab:res}.

We then constructed a 0.1--500.0~GeV light curve for \gsrc\ by binning
the LAT data into 90-day intervals and performing likelihood analysis
to the binned data. To check possible details of the light curve, 
we also constructed a smooth light curve by shifting each 90-day time bin
5\,day forward. The light curves (left panel of Fig.~\ref{fig:lc}) show
that prior to $T_0$ (i.e., MJD~59704.17), no \gr\ emission was 
detected from \sn, but after a flux peak before
the end of 2022, \gsrc\ was detectable possibly up until Oct. 27 2023
(MJD~60244). We performed likelihood analysis to the data from $T_0$
to MJD~60244, and obtained a TS value of 78.1; the other measurements
are provided in Table~\ref{tab:res}. In addition, we also tested the analysis
for the data from $T_0$ to May 10 2023 (MJD~60074; see 
section `Periodicity determination' below) and from 
MJD~60074 to 60244, and obtained TS values of 50.3
and 21.2, respectively. These results confirm
the light-curve results that the \gr\ emission was detectable before 
Oct. 27 2023.
	
By running the tool {\tt gtfindsrc} to $>$1\,GeV data in the $\sim$1.5-year 
time period from $T_0$ to MJD~60244, we obtained the position of \gsrc:
R.A. = 8$^\circ.57$, Decl. = $-$8$^\circ.46$ (J2000.0), with a
3$\sigma$ uncertainty of 0$^{\circ}$.16. \sn\ is within this uncertainty 
region. We updated our source model by using this position for \gsrc. We noted
that the analyses were not sensitive to the two positions, as the light
curves and best-fit results from the binned likelihood analysis  
were nearly identical.

\subsection{Periodicity determination}

We constructed an aperture light curve using the source model file obtained 
above to search and determine the periodicity, where a modified aperture 
photometry (AP) method was employed.
First, since \gsrc~was a faint \gr~source, events within a 6$^{\circ}$ 
radius\footnote{68\% containment angle of the LAT at 0.1\,GeV; https://www.slac.stanford.edu/exp/glast/groups/canda/lat\_Performance.htm}
aperture in 0.1--500\,GeV were selected. To mitigate 
\gr~contamination from the Sun or
the Moon, we excluded the data from time periods when \gsrc~was within 
5$^{\circ}$ of the Sun or the Moon, as determined from {\tt gtmktime}.
Second, time bins of 500\,s each were set, and the LAT exposures of the time 
bins were taken
into account, where the exposures were obtained from the tool {\tt gtexposure}. 
We then calculated the probabilities of the events originating from \gsrc\ 
using {\tt gtsrcprob}, and summed them within each time 
bin \cite{k11,lat+12,cor+19}. The probability values, rather than
the raw event counts, were taken as the counts of the time bins. 
Third, barycentric corrections to the arrival times of the time bins
were applied by using the tool {\tt gtbary}, which considered the motion of 
both the Earth and the Fermi satellite relative to the Solar system's
barycenter.

In performing the timing analysis, we utilized the Lomb-Scargle Periodogram
(LSP) \cite{l76,s82,zk09} to compute the power spectrum from the AP light 
curve. A periodic signal near the 12.4-day optical undulation period was 
easily found from the power spectrum (e.g., the left panel of 
Fig.~\ref{fig:lsp}).
However, given the period was known, we instead conducted
periodicity search analysis in an 11.0- to 14.0-day window.
A periodic signal at $\sim$12.5~day was found from the AP light curve
spanning from $T_0$ to MJD~60244.  Because temporal changes in 
\gr~orbital signals were seen in some binary 
systems \cite{nts+18,cla+21,ccc+22}, 
we examined the variations of this 12.5-day signal
by using AP light curves of different lengths, all starting from $T_0$
with a 30-day time step. The results are shown in the right panel 
of Fig.~\ref{fig:lc}.
The signal peak reached a maximum power value of $\sim$17.8 on May 10 2023
(MJD~60074).
By fitting this power peak with a Gaussian function, the period was determined
to be 12.54$\pm 0.18$\,day, where the uncertainty was the full width at half
maximum of the Gaussian fit (see the right panel of Fig.~\ref{fig:lsp}). 
We note that the 
period is consistent with the optical one within the uncertainty.

\begin{figure}
\begin{center}
\includegraphics[width=0.52\linewidth]{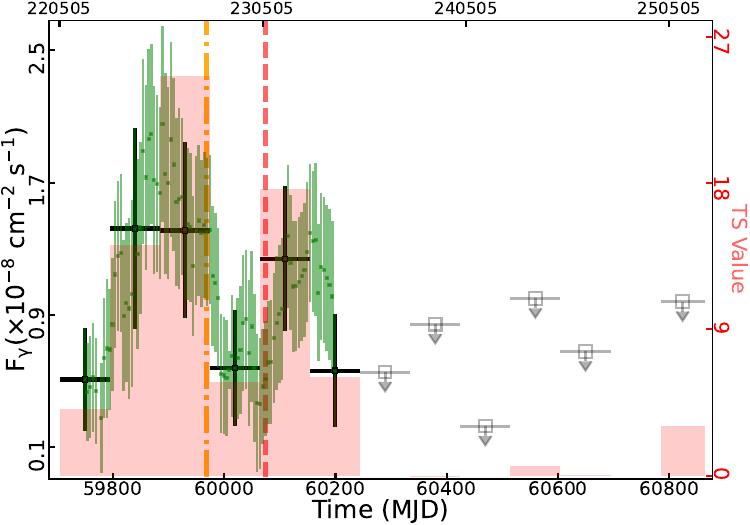}
\includegraphics[width=0.46\linewidth]{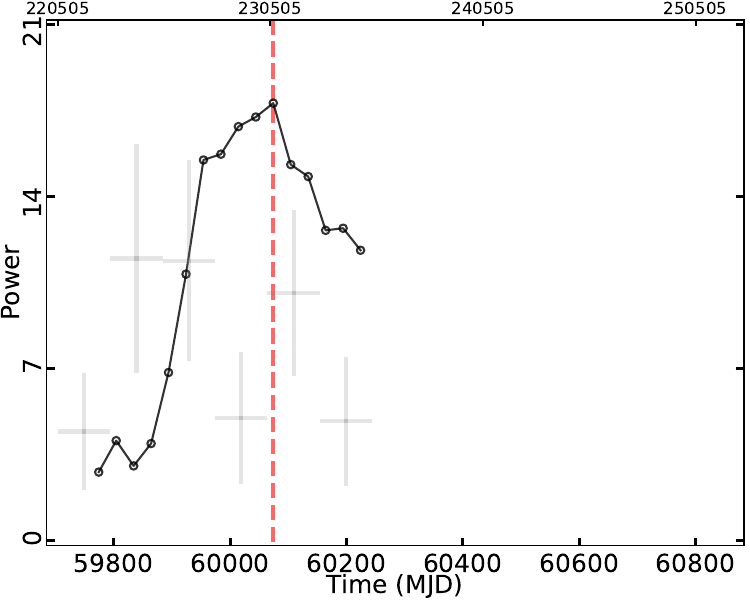}
\end{center}
	\caption{{\it Left:} 90-day binned light curve (black data points)
	of \gsrc\ in 0.1--500\,GeV from $T_0$ to Jul. 7 2025.
	The TS values of the data points are indicated by the red histogram.
	After $\sim$MJD~60244, the source was undetected.
	A smooth 
	light curve (green data points), with each 90-day time bin 
shifted by 5 day forward, is also shown for comparison. The end of the optical
	light curves of \sn\ in ref.~\citenum{che+24} is marked by the
	orange dash-dotted line and the date of May 10 2023 (MJD~60074) is 
	marked by the red dashed line.
	{\it Right:} power peak values of the $\sim$12.5-day signal,
	obtained from LSP analysis of AP light curves of different lengths
	(all starting from $T_0$ with a 30-day time step).
	The maximum value is 17.8 (marked by the red dashed line), when 
	the AP light curve is from $T_0$ to May 10 2023.
\label{fig:lc}}
\end{figure}

In calculating the power spectrum, a frequency range was set from 
$f_{\rm min}$~=~1/14.0~day$^{-1}$ to $f_{\rm max}$~=~1/11.0 day$^{-1}$, 
with a frequency resolution of $\delta f=1/T{_{\rm obs}}$, where 
$T{_{\rm obs}}$ is the time duration of the AP light curve 
(i.e., $\delta f=1/370$\,day$^{-1}$). The number $N$ of independent 
frequencies was
$N=(f_{\rm max} - f_{\rm min})/\delta f \simeq 7$, which could be taken as
the trial factor. Additionally, an oversampling factor of 5 was 
adopted \cite{cor+16,cor+19,ccc+22}. 
Assuming Gaussian white noise, we estimated the probability $p_{_{\rm lsp}}$
of detecting a periodic signal with a power level of 17.8 by chance,
$p_{_{\rm lsp}}\sim1.82\times10^{-8}$. Accounting for the trial factor $N$,
the corresponding false alarm probability (FAP) was 
FAP$ =1-(1-p_{_{\rm lsp}})^N \simeq N\times p_{_{\rm lsp}}=1.27\times10^{-7}$, 
which corresponds to a confidence level of $\sim5.3\sigma$. We note that
because the period can be considered known, we could set analysis with
$N = 1$, which would increase the confidence level to 5.6$\sigma$.

\subsection{Phase-resolved analysis}

Based on the periodicity we obtained above, we divided the LAT data 
in the time range of $T_0$ to MJD~60074 into 10 periodic phase intervals, 
with phase zero set at $T_0$. A folded light curve in 0.1--500~GeV was 
obtained
by performing likelihood analysis to the data in each phase interval.
In this analysis, the flux normalizations of sources within $5^{\circ}$ of
\gsrc\ were allowed to vary, while all other parameters were fixed at the
values in our source model above (from the data of $T_0$ to MJD~60244).
For the obtained flux data points, those with TS $\ge$ 4 were kept; otherwise,
the 95\% flux upper limits were derived and used 
(left panel of Fig.~\ref{fig:plc}). The emission of \gsrc\ was highly variable,
as it was detected in the phase range of 0.0--0.4 (defined as
Ph$_{1-4}$) and the flux peak was at phase 0.2--0.3 (defined as Ph$_3$).
To examine the emission properties, we performed likelihood analysis to
the data in Ph$_{1-4}$ and Ph$_3$, where the parameters were set as the above.
The results are provided in Table~\ref{tab:res}.

It is interesting to compare the \gr\ modulation with the optical undulation.
We used the $g$ and $r$ band data from the Zwicky Transient 
Facility (ZTF) and followed the analysis method in ref.~\citenum{che+24} but
set the period at 12.54\,day. The obtained undulation profiles 
(left panel of Fig.~\ref{fig:plc})
are very similar to those reported in ref.~\citenum{che+24}.

We also obtained the spectra for \gsrc\ in Ph$_{1-4}$ and Ph$_3$ by
dividing the energy range of 0.1--500\,GeV into 10 bins equally 
spaced in logarithmic energy
and performing likelihood analysis to the data in each bin.
The same parameter setup as the above was applied.
For the obtained data points, those with TS $\ge$ 4 were kept; otherwise, 
the 95\% upper limits were derived and used instead.  
For comparison, we also obtained the spectrum in the 
$T_0$ to MJD~60244 time period
(i.e., the total data for \gsrc). The same
analysis was performed. No significant differences among the spectra
could be determined (see the right panel of Fig.~\ref{fig:plc} and 
Table~\ref{tab:res}).

\end{methods}


\section*{Data Availability}
The data results that support the findings of this study are available from
the corresponding authors upon reasonable request.

\bibliography{exrb}



\begin{addendum}
 \item 
This work is supported in part by the National Natural Science Foundation
of China under grant Nos. 12233006, 12163006, and 12273033,
	and the joint foundation of
Department of Science and Technology of Yunnan Province
and Yunnan University (grant No. 202201BF070001-020).
P.Z. acknowledges the support by the Xingdian Talent Support
Plan–Youth Project.  

\item[Contributions]
P.Z. led the analysis and made the initial discovery.
Z.W. provided explanations and wrote most of the text. S.J. aided 
the explanations. All authors discussed the results and manuscript.

 \item[Competing Interests] The authors declare that they have no
competing financial interests.
 \item[Correspondence] Correspondence and requests for materials
should be addressed to P.Z.~(email: zhanngpengfei@ynu.edu.cn) and
Z.W.~(email: wangzx20@ynu.edu.cn).
\end{addendum}


\begin{table*}
\begin{center}
\caption{Results of likelihood analyses}
\begin{tabular}{lccccc}
\hline\hline
Time range &  TS & $\Gamma$ & $F\rm_{ph}$ & $F\rm_{en}$ & $L_\gamma$ \\
\hline
20080804--20250707 & 22.6 & 2.50$\pm$0.20 & 2.68$\pm$0.80 & 1.26$\pm$0.41 & 0.76$\pm$0.25 \\
20220505--20231027 & 78.1 & 2.15$\pm$0.12 & 9.5$\pm$2.4 & 8.5$\pm$1.8 & 5.1$\pm$1.1 \\
20230510--20231027 & 21.2 & 2.10$\pm$0.20 & 9.3$\pm$3.8 & 9.3$\pm$3.8 & 5.7$\pm$2.3 \\
20220505--20230510 & 50.3 & 2.25$\pm$0.16 & 10.9$\pm$3.3 & 7.7$\pm$1.7 & 4.6$\pm$1.0 \\
\hline
Phase & \multicolumn{5}{c}{Phase-resolved analysis}  \\
\hline
Ph$_{1-4}$ & 81.5 & 2.23$\pm$0.13 & 26.1$\pm$6.0 & 19.1$\pm$4.1 & 11.6$\pm$2.5\\
Ph$_{3}$ & 34.5 & 2.26$\pm$0.19 & 42$\pm$14 & 29.3$\pm$7.4 & 17.7$\pm$4.5\\
\hline
\end{tabular}
\label{tab:res}
\end{center}
{\bf Notes. }{The integrated photon flux ($F_{\rm ph}$) is given in units of
              $10^{-9}$~photons~cm$^{-2}$~s$^{-1}$, the integrated 
	      energy flux ($F_{\rm en}$) in units of 
	      $10^{-12}$~erg~cm$^{-2}$~s$^{-1}$, and
              the \gr~luminosity ($L_{\gamma}$) at a distance of 22.5\,Mpc 
	      in units of $10^{41}$~erg~s$^{-1}$}.
\end{table*}

\end{document}